\documentclass[twocolumn, prb, showpacs]{revtex4}
\usepackage{amssymb}
\usepackage{graphicx}
\usepackage{dcolumn}
\usepackage{bm}
\usepackage{amsmath}
\begin{document}
\title{Two-dimensional spin-1 frustrated Heisenberg model with valence-bond ground states}
\author{Zi Cai$^{1}$, Shu Chen$^{1,\ast}$, Supeng Kou$^{2}$, and Yupeng
Wang$^{1}$}
\address{ $^1$Beijing National Laboratory for
Condensed Matter Physics, Institute of Physics, Chinese Academy of
Sciences, Beijing 100080, P. R. China\\
$^2$Department of physics, Beijing Normal University, Beijing
100875, P. R. China }

\date{Received \today }

\begin{abstract}
In this paper, we propose a method to understand the nature for the
quantum disorder phase of the two-dimensional (2D) high spin
frustrated model. The ground state and excitation properties of a
fully frustrated 2D spin-1 model are studied based on a model whose
groundstate can be found exactly. By analogy to the pseudo-potential
approach in the fractional quantum Hall effect, we conclude that the
ground states of the fully frustrated spin-1 model are doubly
degenerate valance bond solid (VBS) states along the horizontal or
vertical direction of the square lattice. We also find that the
ground state could be characterized by a nonzero string order, which
rarely happened in the 2D case. The method that we used reveals the
connection between the fractional quantum hall effect and the
frustrated 2D antiferromagnetism system. The VBS states capture the
main character of the disordered phase in the 2D spin-1 frustrated
system, and can be verified by a numerical method.
\end{abstract}


\maketitle
\section{Introduction}
The 2D quantum frustrated magnets have attracted considerable
attention in the past years \cite{Misguis} because they are believed
to be promising candidates for realizing exotic spin liquid states.
One of the most studied 2D frustrated models is the
antiferromagnetic Heisenberg model on a square lattice with the
first and second nearest neighbor interactions $J_1$ and $J_2$,
\begin{equation}
H=J_1\sum_{\langle \mathbf{i},\mathbf{j}\rangle
}\mathbf{S}_{\mathbf{i} }\cdot
\mathbf{S}_{\mathbf{j}}+J_2\sum_{\langle \langle
\mathbf{i},\mathbf{j} \rangle \rangle
}\mathbf{S}_{\mathbf{_i}}\cdot \mathbf{S}_{\mathbf{_j}}\,
\end{equation}
where $\langle \mathbf{i},\mathbf{j}\rangle $ and $\langle \langle
\mathbf{i} ,\mathbf{j}\rangle \rangle $ represent pairs of the
nearest and next-nearest neighbors respectively. The general ground
state properties of this model have been investigated in various
methods \cite{2,3,4,Read,5}. It has been well accepted that the
$J_1-J_2$ model exhibits a quantum phase transition from a
magnetically ordered N\'{e}el phase at small $J_2/J_1$ to a spin
disordered phase in the intermediate parameter region. Despite the
success in predicting the order to disorder transition, the theory
based on mapping onto the non-linear $\sigma $ model and symmetry
analysis provides no clear information about the nature of the
disordered phase. For the 2D spin-$1/2$ $J_1-J_2$ model, there have
been a lot of works focusing on its nonmagnetic phase, but no
definite conclusion has been drawn on the controversial question
whether the disordered phase is a spin liquid without symmetry
breaking or a valence bond crystal with broken spatial symmetry
\cite{Capriotti01,Sushkov}.

Most researches in this field  focus on the spin-$1/2$ system
because of its possible correlation with the High-Tc
superconductivity. However, in recent years ultracold atomic
system has provided an ideal playground to experimentally
investigate the high-spin strongly correlated system. Several
proposals have been provided to realize the spin-1 lattice
model\cite{Cirac}. Considering the rapid development in this
field, we expect that it is possible for  the spin-$1$ $J_1-J_2$
model to be realized experimentally in the future. However, in
comparison with the spin-$1/2$ system, little knowledge is known
about the disordered phase of the 2D spin-1 system except  the
possible double degeneracy of the ground state predicted by the
field theory\cite{Read}.

In this paper we first review some general results of the
disordered phase in the frustrated model, then we investigate the
ground state properties of a fully frustrated spin-$1$ $J_1-J_2$
model with $J_2/J_1=0.5$ by the pseudo-potential approach. Our
method can be generalized to deal with other systems such as the
frustrated honeycomb model and system with higher spin.

The general property for the 2D spin 1 frustrated model was
investigated in the scope of the (2+1)-dimensional nonlinear $\sigma
$ model\cite{3,4}, which is considered as the continuum limit of the
SU(N) antiferromagnet \cite{Read,7}. Though this model was derived
in the semiclassical (large S) limit and no frustrated condition, it
is still valid to describe the condition of small S and frustrated
case\cite{3,4}.The effective action of a 2D spin-1 $J_1-J_2$ model
is
\begin{equation}
S_{\rm eff} = \frac{1}{2g_0}\int \! d\tau d^2{\bf r} \biggl\{
\frac{1}{c} (\partial_{\tau}{\bf m})^2 + c (\nabla{\bf m})^2
\biggl\}+S_{\rm B}
\end{equation}
where $S_B=iS\sum_{\mathbf{R}}\eta _R\oint dm_R\cdot
\mathbf{A_R}$, $g_0= \frac{\sqrt{8}a}{S\sqrt{1-2\alpha }}$ and
$c=\sqrt{8\left( 1-2\alpha \right) }J_1Sa$ with $\alpha
={J_2}/{J_1}$. This is precisely the O(3) non-linear $ \sigma $
model in the (2+1) dimensions with a residual Berry phase
$(S_{\mathrm{B}})$. Via this model, renormalization group (RG)
analysis predicted that there is a transition from the N\'{e}el
state to a disordered state and the transition point for the
spin-1 model is $\alpha _c=0.46$, larger than that in spin-1/2
case. The Berry phase plays a crucial role in determining the
degeneracy and symmetry of the ground state in the disordered
phase\cite {7,Read}. For the spin-1 model, the degeneracy in the
disordered phase is 2 and there is a spontaneous symmetry
breaking. We will see below  that the symmetry of our trial ground
states completely agrees with that predicted by the general theory
above.

The analogy between the Heisenberg model and the fractional
quantum Hall effect (FQHE) was first introduced by Arovas {\it et
al.} \cite{Arovas} in the one-dimensional (1D) case. They
decomposed the spin-1 antiferromagnetic Heisenberg Hamiltonian as:
\begin{equation}
H_1=\sum_{i} \mathbf{S_i}\cdot \mathbf{S_{i+1}} = \sum_i [3
P^2(i,i+1)+P^1(i,i+1)-2]
\end{equation}
while they treat the projection operator $P^1(i,i+1)$ over spin-1
states associated with two consecutive sites as a perturbation and
consider the Affleck-Kennedy-Lieb-Tasaki (AKLT) state\cite{AKLT},
which is the exact ground state of the projection operator $
3P^2(i,i+1)$ over spin-2 states, as a trial ground state of the
spin-1 Heisenberg chain. The AKLT state and the real ground state
of the spin-1 Heisenberg chain share a lot of important
properties, for example both of them have an energy gap and the
two point correlation functions decay exponentially. They are in
the same universal class and characterized by a hidden order
parameter called ``string parameter".\cite{17} Numerical results
show that the difference between the ground state energies of
these two states is within $5\%$. In this work, we extended their
method to the 2D fully frustrated model to study the properties of
the groundstate and the elementary excitation.

For the fully frustrated model with $J_2/J_1=0.5$, we can rewrite
the Hamiltonian Eq.(1) as:
\begin{equation}
H=\sum_\alpha H_\alpha \quad with\quad H_\alpha
=J_2\sum_{\mathbf{{i},{j}\in \alpha
}}\mathbf{S}_{\mathbf{_i}}\cdot {\bf S}_{\mathbf{_j}},
\label{J1J2}
\end{equation}
where the index $\alpha $ denotes the site of the dual lattice corresponding
to the $\alpha $th plaquette. Each bond interaction between the two spins in a
plaquette is equal. For convenience, we expand this Hamiltonian by the
projection operators of total spin in a plaquette:
\begin{equation}
H_\alpha /J_2=\sum_{S=0}^4C_S\mathbf{P_\alpha ^S}-4
\end{equation}
with $C_s=S(S+1)/2$ and further decompose it as the summation of two parts:
\[
H_\alpha =H_\alpha ^0+H_\alpha ^1,
\]
with
\begin{eqnarray*}
H_\alpha ^0/J_2&=&10\mathbf{P_\alpha ^4}+6\mathbf{P_\alpha ^3}-4,\\
H_\alpha ^1/J_2&=&3\mathbf{P_\alpha ^2}+\mathbf{P_\alpha ^1}.
\end{eqnarray*}
The operator $ P_\alpha ^S$ projects the spin state of the $\alpha
$th plaquette onto the subspace with total spin $S$. One of the
reasons for dividing the original Hamiltonian into two parts is
that the model $H^0=\sum_\alpha H_\alpha ^0$ is quasi-exactly
solvable in the sense that its ground state can be obtained
exactly. Furthermore, we observe that the coefficient $C_S$
decreases rapidly as $S$ descends. Therefore it is reasonable to
apply the pseudo-potential method originally used in the FQHE
\cite{Haldane83}. Using this method, we first classify all the
ground states of $H^0$ and treat $H^1=\sum_\alpha H_\alpha ^1$ as
a perturbation. The ground state of $H^0$ is shown to be highly
degenerate and belongs to two kinds of states: the dimer states
and the valence bond solid states. However the huge degeneracy of
the ground states is lifted when $H^1$ is introduced. Finally our
result shows that the ground states of the fully frustrated spin-1
model are doubly degenerate and approximately described by the
decoupled VBS states along the horizontal or vertical direction of
the square lattice. We believe that our work would inspire the
interest in the spin-1 frustrated model.

\section{The spin-1 model with exact ground states}

Now we start by considering a spin-1 Hamiltonian on a square
lattice:
\begin{equation}
H_p=\sum_\alpha \left( c_4\mathbf{P_\alpha ^4}+c_3\mathbf{P_\alpha ^3}%
\right) ,  \label{Hp}
\end{equation}
where $H_p$ is an arbitrary linear composition of projection operators $%
P_\alpha ^4$ and $P_\alpha ^3$ with $c_4,c_3\geq 0$. Obviously, $H^0$ is
just a special case of the above Hamiltonian. Since $H_p$ is positive
semidefinite, the state with the total spin of each plaquette $S_\alpha
^T\leq 2$ is the exact ground states for $H_p$. Such a condition could be
satisfied by two kinds of states: \textit{the dimer states} and \textit{the
valence bond solid states}.

\begin{figure}[htb]
\includegraphics[width=3.4in]
{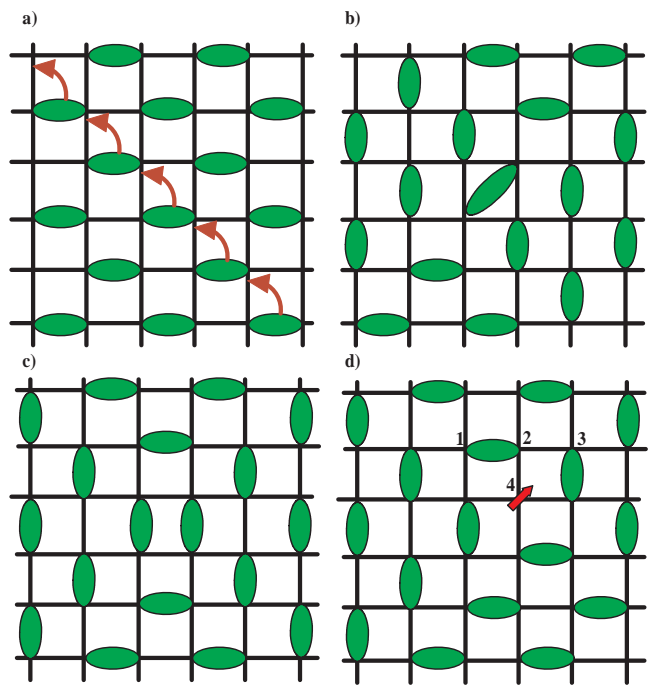} \caption{(Color online) Different dimer ground
states of $H_p$. The green (dark) ellipsoids represent singlet
dimers formed by two $S=1$ spins.} \label{fig1}
\end{figure}

The dimer states are the states with each plaquette possessing a
dimer, which is a spin singlet formed by two neighboring spins:
$\frac 1{\sqrt{3} }(|1,-1 \rangle +|-1,1 \rangle -|00\rangle )$.
This family of states are illustrated in Fig.1 and are very
similar to the dimer ground states in the spin-1/2
case\cite{Batista} except the singlet dimers are formed by pairs
of spin with $S=1$ and also the defect spin in Fig. 1(d) has
$S=1$. As stated by Batista {\it et al.}, the states in Fig.1
(a-d) represent all the possible configurations which satisfy the
constrain that every plaquette possesses at least one
dimer\cite{Batista}. For the configuration in Fig.1(a), rotating
any array of dimers along a diagonal direction by $\pi /2$ , we
will get a new degenerate state of this model, as stated in
\cite{Batista}. The degeneracy of this kind of ground state is
proportional to $2^{\sqrt{N}+1}$, where $N$ is the total number of
the sites. States in Fig.1(b-d) represent the configurations with
one "defect" and it is possible to have a localized $S=1$ spin as
illustrated in Fig. 1(d). Translating or rotating these kinds of
states around the central defect gives a huge amount of degeneracy
proportional to N, the total number of the sites. Configurations
in Fig. 1(a-d) and their corresponding degeneracy states provide a
full configuration subspaces that satisfy the constraint that each
plaquette has at least one singlet.

\begin{figure}[tbh]
\includegraphics[width=3.4in]
{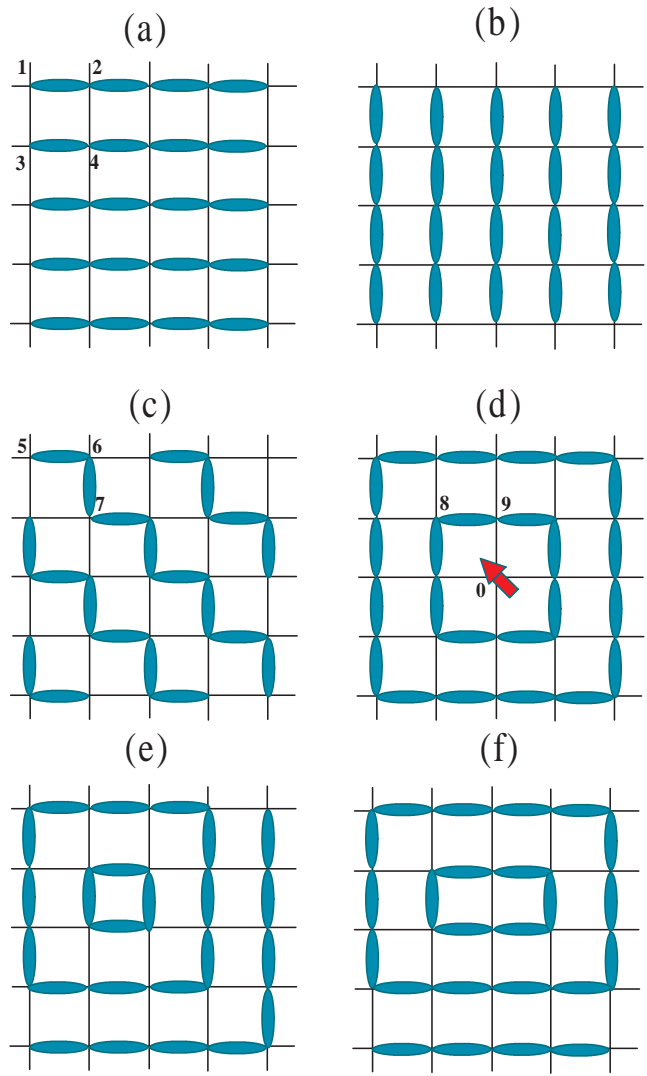} \caption{(Color online) Different AKLT ground states
of $H_p$. The blue (dark) ellipsoids represent singlet states
formed by two $S=1/2$ pseudospins.} \label{fig2}
\end{figure}

For the spin-1 system, however, there may be another family of
ground state totally different from that of the spin-1/2 case: the
VBS state, or the AKLT state \cite{AKLT}. An $S=1$ spin in one
site could be considered as a symmetric composition of two $S=1/2$
spins and the AKLT state is formed by the valence bond, which is
the singlet state of two neighboring $S=1/2$ spins. If there is a
global state ensuring that each plaquette has at least two valence
bonds, as illustrated in Fig.2(a-f), such a state is also a ground
state of $H_p$ due to $S_\alpha ^T\leq 2$. Obviously, this kind of
ground state is also highly degenerate. The lattice we considered
is under the open boundary condition with the total site of
lattice $N\rightarrow\infty$. Configurations in Fig.2(a-c) are
composed by many 1D AKLT chains with infinite length. Fig.2(d-e)
represent the configurations composed by the closed AKLT loop and
a defect in the center. Other configurations not listed here which
satisfy the same constraint are similar to Fig.2(e,f), but with a
defect of a different shape. We could estimate that the number of
this kind of degeneracy is proportional to $N$ by considering the
possible positions of the central defect in  Fig.2(d-f).

Now we focus on the excitation of the second family. The
elementary excitation of the AKLT chain is the magnon excitation
with a Haldane gap of 0.35\cite{Arovas}, in which a singlet dimer
formed by two neighboring $S=1/2$ spins is excited to a triplet
state, as illustrated in Fig.3(a). Unlike the dimer case, these
two 1/2 spinons could not be separated without increasing the
energy proportional to the distance between them, so this kind of
excitation is confined. However, the triplet state itself, which
is formed by two binding spinons, could propagate along the AKLT
chain freely, as shown in Fig.3(b).

It is noted that a similar method has been used to construct the
exact ground state of the well known 1D AKLT model \cite{AKLT} and
the 2D spin-1/2 model with additional cyclic
exchanges\cite{Batista}. The application of such a method has made
a lot of success in understanding the nature for spin
systems\cite{AKLT,Batista,Majumdar,Chen,Nussinov}.
\begin{figure}[tbh]
\includegraphics[width=3.4in]
{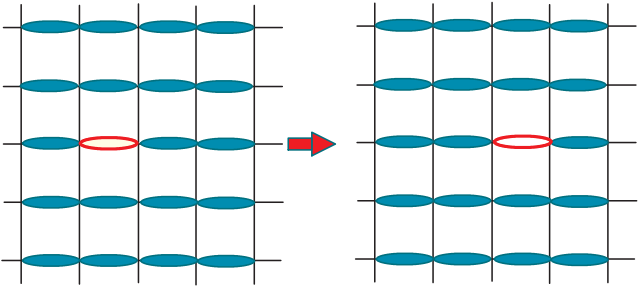} \caption{(Color online) (a) One singlet is excited
to a triplet (open ellipsoid). (b) The excitation could propagate
along the AKLT chain freely. The blue (dark) ellipsoids represent
singlets formed by two pseudospins with $S=1/2$. } \label{fig3}
\end{figure}

\section{The ground state of the $J_1-J_2$ model with full
frustration}

Since the Hamiltonian Eq.(\ref{Hp}) is somehow an artificial
model, how could we expect that it is related to the fully
frustrated $J_1-J_2$ model? We will clarify this problem below
using the analogy to the pseudopotential approach in the FQHE
\cite{Haldane83}, where, restricted within the first Landau level,
the interaction can be expanded by relative angular momentum
projection operators: $v(r_i-r_j)=\sum_{k=0}^\infty v_kP_k(ij)$
with $P_k(ij) $ the projection operator on states with angular
momentum $k$. If a potential satisfies $v_k$=0 when $k\geq m$, the
Laughlin state is the exact ground state for this model. The
success of this method depends on the difference between the
coefficient $v_k$ of $k=m-2$ and $k=m$\cite{Haldane83}.

Now we treat the 2D $J_1-J_2$ Hamiltonian in the spirit of the
pseudopotential approach  and consider $ H^1_\alpha$ as
perturbation of $H^0_\alpha$, just as Arovas et. al did in the 1D
case. Since the ground state of $H^0=\sum_{\alpha}H^0_{\alpha}$
has been analyzed above, all the degenerate ground states of
Hamiltonian (\ref{Hp}) may be a possible candidate of the trial
ground state for the Hamiltonian (\ref{J1J2}). However, the
perturbation $H^1=\sum_{\alpha}H^1_\alpha$ lifts the high
degeneracy and chooses some states as the best trial ground state.

Next we will calculate the energy expectation under those
degenerate states. The lower the expectation value is, the closer
the state is to the real ground state of (\ref{J1J2}). We notice
that our lattice is constructed by the bonds connecting one site
with its nearest neighborhood or the next nearest ones. To
calculate the average energy  of various candidate trial
groundstates, we must first calculate the energy expectation of
each bond. For the dimer states as shown in Fig.1, all the
configurations in Fig.1 (a-d) are constructed by three kinds of
bonds. Take the Fig.1 (d) for an example: the first kind is
represented by bond $12$, which is covered by a single dimer. Bond
$23$ represents the bond between two different dimers, and the
third kind (bond$24$) is the one between a dimer state and a
single spin with $S=1$. The energy expectations for these three
kinds of bonds are
\begin{eqnarray}
\langle\mathbf{S}_{\mathbf{_1}} \cdot {\bf S}_{\mathbf{_2}
}\rangle_{dimer} &=&-2,\\
\langle\mathbf{S}_{\mathbf{_2}} \cdot {\bf S}_{\mathbf{_3}
}\rangle_{dimer} &=& \langle\mathbf{S}_{\mathbf{_2}} \cdot {\bf
S}_{\mathbf{_4} }\rangle_{dimer} = 0, \nonumber
\end{eqnarray}
respectively. It is ready to calculate the average energy per
plaquette in the thermodynamic limit:
\[
\quad \langle E\rangle_{dimer}=-2 J_2 ,
\]
which is the same for all the
configurations in the Fig.1 (a-d).

Now we turn to the VBS states in Fig.2. Similar to the method
above, we can classify all the bonds which construct the
configurations in Fig.2 into five classes: (1) Bond $12$ or $34$
in Fig.2(a), which is covered by a valence bond formed by the
singlet of two $1/2$ spinons in a AKLT chain with infinite length;
(2) Bond $57$ in Fig.2(c) is the bond connecting two next nearest
sites in an AKLT chain; (3) Bond $13$ or $24$ in Fig.2(a) which
connects different AKLT chains; (4) Bond $89$ in Fig.2(d) is
similar to the first one, but the bond is covered by a valence
bond in a closed AKLT loop with finite length; (5) Bond $90$ in
Fig.2(d) is the bond between one site in an AKLT chain and another
site occupied by a single spin with $S=1$.

Now we calculate the average energy under all these candidate
trial ground states in Fig.2 to find which one has the lowest
energy. We define $\mathbf{T}_{\pi/2}$ as an operator to rotate
the system by $\pi/2$ around one site in the lattice. So we can
see that $\Psi _{AKLT\Vert }^a$=$\mathbf{T}_{\pi/2}$$\Psi
_{AKLT\Vert }^b$ , and the Hamiltonian in Eq.(1) is invariant
under the operator $\mathbf{T}_{\pi/2}$. So the states in Fig.2(a)
and (b)
are  degenerate. We denote these states as $%
\Psi_{AKLT\|}$. In a similar way, we can get
\[
\langle\mathbf{S}_{\mathbf{_1} } \cdot
\mathbf{S}_{\mathbf{_2}}\rangle_{AKLT\|}=\langle\mathbf{S}_{\mathbf{
_3}} \cdot \mathbf{S}_{\mathbf{_4}}\rangle_{AKLT\|}=-4/3
\]
while other bonds all contribute zero to the average energy. It is
clear that we have
\[
\langle E\rangle_{AKLT\|}=-\frac{8}{3}J_2 ,
\]
which is smaller than $\langle E\rangle_{dimer}$. If we consider
the states like Fig.2(c), where there are many right angles along
the AKLT chain, we find that the average energy under these states
denoted by $\Psi_{AKLT\perp}$ is always higher than that of
$\Psi_{AKLT\|}$. In the AKLT state, there is a short range
correlation \cite{AKLT}, and thus we get
\[
\langle\mathbf{S}_{ \mathbf{_5}} \cdot
\mathbf{S}_{\mathbf{_7}}\rangle=4/9 ,
\]
whereas we have $\langle\mathbf{S}_{\mathbf{_5}} \cdot
\mathbf{S}_{\mathbf{_6 }}\rangle=\langle\mathbf{S}_{\mathbf{_6}}
\cdot \mathbf{S}_{\mathbf{_7}}\rangle=-4/3$. So this kind of bonds
in the corner of the AKLT chain would always raise the average
energy of the system. We can find in Fig.2(d)
$\langle\mathbf{S}_{\mathbf{_9}} \cdot
\mathbf{S}_{\mathbf{_0}}\rangle=0 $  So the energy of
configurations corresponding to Fig.2(d-f) and other
configurations not listed here is always higher than the energy of
$ \Psi _{AKLT\Vert }^a$ and $\Psi _{AKLT\Vert }^b$ for two
reasons: the appearance of right angles along an AKLT chain as
well as the fact that the average value of
$\mathbf{S}_{\mathbf{_i} } \cdot \mathbf{S}_{\mathbf{_i+1}}$ is
lower in an AKLT chain with infinite length than that in a closed
AKLT loop with finite length, {\it i.e.},
$\langle\mathbf{S}_{\mathbf{_1} } \cdot
\mathbf{S}_{\mathbf{_2}}\rangle_{AKLT\|}<\langle\mathbf{S}_{\mathbf{
_8}} \cdot \mathbf{S}_{\mathbf{_9}}\rangle_{AKLT\Box}$. The proof
of this fact has been shown in the Appendix.

As we showed above, the perturbation lifts the degeneracy and
chooses state $ \Psi _{AKLT\Vert }^a$ and $\Psi _{AKLT\Vert }^b$
(shown in Fig.2 (a, b)) as trial ground states. Apparently, they are
doubly degenerate. It is natural to ask how close they are to the
real ground states. Here we give an estimation. We can expand $\Psi
_{AKLT\Vert }^a$ in the spirit of the pseudo-potential approach:
$\Psi _{AKLT\Vert }^a$=$a\Psi _2+b\Psi _1$, where $\Psi _\beta
=\sum_\alpha P_\alpha ^\beta \Psi _{AKLT\Vert }^a$ ($\beta =1,2$)
and $a^2+b^2=1$. The reason why there is no component of $\Psi _0$
is that if there exists a state in which the total spin of every
plaquette is zero, the energy for every plaquette gets to its
minimum simultaneously which is impossible because the wave function
would not be self-consistent. Observing that $\langle
\mathbf{S}_{total}^2\rangle _{AKLT\Vert }=6a+2b=8/3$,  we can get
the coefficient $b=0.994$, which means $\langle \Psi _{AKLT\Vert
}^a|\Psi _1\rangle =0.994$. Based on this result, we can reconfirm
the validity of the perturbation used above. Because $\Psi _1$ is
the dominant part in $\Psi _{AKLT\Vert }^a$, the operator
$\mathbf{P_\alpha ^1}$ in the perturbation $H^1$ plays a more
important role than $\mathbf{P_\alpha ^2}$. Furthermore since the
coefficient of $\mathbf{P_\alpha ^1}$ is smaller than that of
$\mathbf{P_\alpha ^2}$, the effect of the perturbation to the
unperturbed state is even smaller. Therefore, we believe that this
trial state should be a good approximation to the ground state of
Hamiltonian (\ref {J1J2}).

Nevertheless, it seems that there is still a question to be
clarified. Since the trial ground states $\Psi^{a}_{AKLT\|}$ and
$\Psi^{b}_{AKLT\|}$ are degenerate, we should use the degenerate
perturbation rather than that used above, which means that a
linear composition of these two degenerate states could further
lower the energy and forms a new ground state, just as in the
resonant valence bond theory. We shall show that it is not true in
this case. As we know, whether the degenerate perturbation works
depends on the non-diagonal term
$\langle\Psi^{a}_{AKLT\|}|H|\Psi^{b}_{AKLT\|}\rangle$ vanishing or
not. For an $N \times N$ square lattice model with periodical
boundary condition, the wavefunction corresponding to Fig. 2(a)
can be represented as
\begin{equation}
\Psi^{a}_{AKLT\|}=\Psi^{a_1}_{AKLT\|}\otimes
\Psi^{a_2}_{AKLT\|}...\otimes\Psi^{a_N}_{AKLT\|},
\label{Wavefuc_a}
\end{equation}
with $\Psi^{a_n}_{AKLT\|}$ denoting the wavefunction of the $n-$th
periodical 1D AKLT chain. We estimate $\langle
\Psi^{a}_{AKLT\|}|H|\Psi^{b}_{AKLT\|}\rangle \sim N^2 2^{-N^2}$,
which decreases extremely rapidly when N increases. So we can
safely draw the conclusion that in the thermodynamic limit, the
non-diagonal term is zero and the degenerate perturbation and
non-degenerate perturbation give the same result.

\section{Excitation}

We can also get some results of the low energy excitation based on
our trial ground state wave function. Taken the
$\Psi^{a}_{AKLT\|}$ for example, the basic picture of the low
energy excitation is illustrated by Fig.3 where a singlet in an
AKLT chain is excited to a triplet and propagates along the chain.
Now we calculate some quantitative results of this picture using a
variational technique. Since our trial ground state wave function
could be written as the production of many parallel 1D AKLT chains
and the correlation function between different AKLT chains is
zero, the low energy excitation is similar to the 1D
case\cite{Sen,Arovas}. To find the variational wave function of
the low energy excitation, we use the matrix product
state.\cite{Zittartz} At a site n, we define the matrix
\begin{equation}
M_n = \left(
\begin{array}{cc}
\sqrt{1/3} ~|0 \rangle_n & \sqrt{2/3} ~|-1 \rangle_n \\
- \sqrt{2/3} ~|1 \rangle_n & - \sqrt{1/3} ~|0 \rangle_n \end{array}
\right) ~. \label{mn}
\end{equation}
In terms of the matrix product, the ground state of the $i-$th
AKLT chain could be represented as
\begin{equation}
|\Psi^{a_i}_{AKLT\|}\rangle= \prod_{n=1}^N M_n^{i} .
\end{equation}
and the overall ground state is given by Eq.(\ref{Wavefuc_a}).

The excited state as shown in Fig.3(a) could be represented as
\begin{equation}
\Phi^{a}_{AKLT\|}({\bf n})=\Psi^{a_1}_{AKLT\|}\otimes
...\Phi^{a_i}_{AKLT\|}(n)...\otimes\Psi^{a_N}_{AKLT\|}
\end{equation}
where  $\Phi^{a_i}_{AKLT\|}(n)$ denotes the state with the $n-$th
singlet of the $i-$th AKLT chain being excited to a triplet. For
brevity, we write $|\Phi^{a_i}_{AKLT\|}(n)\rangle$ as $|n\rangle$
which can be represented in terms of the matrix product as
\begin{eqnarray}
|n\rangle &=& \prod_{m=1}^{n} ~M_m \otimes \left(
\begin{array}{cc}
\sqrt{1/3} ~|0 \rangle_{n+1} & \sqrt{2/3} ~|-1 \rangle_{n+1}\\
 \sqrt{2/3} ~|1 \rangle_{n+1} &  \sqrt{1/3} ~|0 \rangle_{n+1} \end{array}
\right) \nonumber\\
& & \otimes \prod_{m=n+2}^{N} ~M_m ~.\label{}
\end{eqnarray}
This is a state with $S_{tot}^z =0$. After some algebra, we find
that
\begin{eqnarray}
\langle m|n\rangle &=& \delta_{m,n} +\frac{1}{9}(\delta_{m,n-1} +
\delta_{m,n+1}) \\
 \langle m|H|n\rangle &=& \frac{8J_1}{9}\delta_{m,n}
-\frac{J_1}{9}(\delta_{m,n-1} + \delta_{m,n+1})
\end{eqnarray}
where $H$ is the Hamiltonian (4) or (1) with $J_1=2J_2$. A
momentum eigenstate is thus defined as
\begin{equation}
 |\mathbf{k}\rangle ~=~ \sum_\mathbf{n} ~e^{i\mathbf{k}\cdot\mathbf{n}}
 ~|\Phi^{a}_{AKLT\|}({\bf n})\rangle,
\end{equation}
which satisfies
\begin{eqnarray}
\langle\mathbf{k}|\mathbf{k}\rangle &=& (1+\frac{2}{9}cos k_x)N^2, \\
\langle\mathbf{k}|H|\mathbf{k}\rangle &=& (\frac{8}{9}-\frac{2}{9}
cos k_x)N^2J_1.
\end{eqnarray}
So the variational energy is
\begin{equation}
E_{var}(\mathbf{k})=\frac{8-2cos k_x}{9+2cos k_x} J_1,
\end{equation}
where we have made an energy shift of $-\frac{8}{3}J_2 N^2$ in the
above calculation.

\section{Discussion and generalization}

The string order was first observed in the spin-1 AKLT chain
\cite{17}. It is believed that even when the Hamiltonian deviates
from the exact AKLT point, the string order did not
vanish\cite{18}. However, the nonzero string correlator  was
rarely observed in the 2D model, the only exception that we have
known up to now is  the Wen-Kitaev Model\cite{Wen}, which is a 2D
exactly solvable model with a string correlator attaining finite
value. In our trail ground state, since there is a strong evidence
that the 2D ground state is decoupled into several 1D AKLT chains,
the order parameter \cite{17} proposed to characterize the 1D VBS
state should also be valid in our 2D problem. We believe that the
string order parameter:
\begin{equation}
\O^{\alpha}_{string}=-\lim_{|k-l|\rightharpoonup\infty}
\langle\Psi_0|S^{\alpha}_{ik}exp[i\pi\Sigma^{l-1}_{j=k+1}
S^{\alpha}_{ij}]S^{\alpha}_{il}|\Psi_0\rangle\
\end{equation}
should get its maximum in the point $J_2/J_1$=1/2, which is still to
be verified by numerical methods.

Now we discuss the case when the coupling coefficient $J_2/J_1$
deviates from $1/2$. For the general $J_1-J_2$ model, the RG
analysis based on the nonlinear $\sigma $ model can show that
$J_2/J_1 =1/2$ represents a stable fixed point in RG flow in the
parameter space \cite{3}, so we believe that our trial ground
states represent a universal class  and capture the basic property
of the spin disordered phase of the spin-1 $J_1-J_2$ model.

\section{Conclusion}
Based on a solvable spin-1 model with dimer-type and
valence-bond-solid-type ground states, we studied the fully
frustrated spin-$1$ $J_1-J_2$ model and proposed a possible trial
ground state at the point $J_2/J_1=0.5$, which has been seldom
studied before. Such ground states are doubly degenerate and
approximately described by the decoupled AKLT states along the
horizontal or vertical direction of the square lattice. We believe
that this trial state captures the main character of the
disordered phase in the 2D spin-1 frustrated system, and can be
detected by numerical methods. In addition, the pseudo-potential
method used here is not only restricted to square lattice and the
spin-1 case, but also to higher spin or other lattice cases as
long as the Hamiltonian could be expanded to the summation of
projection operators. Taking the spin-1 Honeycomb lattice for an
instance, the state with $S_\alpha ^{total}\leq 2$ for each
hexagon is close to the real ground state of the $J_1-J_2-J_3$
model in the Honeycomb lattice, even closer than the case of the
square lattice because the perturbation takes a smaller portion in
this Hamiltonian than that in the square lattice $J_1-J_2$ model.
So our method offers a potential framework to explore quantum
exotic state in spin systems.

This work is supported in part by NSF of China under Grant No.
10574150 and programs of Chinese Academy of Sciences.

\appendix
\section{}

In this appendix ,we will prove the fact that the average value of
$\mathbf{S}_{\mathbf{_i} } \cdot \mathbf{S}_{\mathbf{_i+1}}$ is
lower in a AKLT chain with infinite length than that in a closed
AKLT loop with finite length. Our proof is based on some results
in Ref \cite{AKLT}.

Let $(+),(0)$,and $(-)$ denote the orthonormal basis for spin 1
consisting of eigenstates of $S^z$ with eigenvalues $+1,0$,and
$-1$,respectively and
\begin{equation}
 \psi_{11}=\sqrt{2}(+),\quad \psi_{12}=\psi_{21}=(0),\quad
\psi_{22}=\sqrt{2}(-) .
\end{equation}
We denote the wavefunction of a single AKLT chain with length L as
\begin{equation}
\Omega_{\alpha\beta}=\psi_{\alpha\beta_1}\bigotimes\psi_{\alpha_2\beta_2}\cdots\psi_{\alpha_
L\beta}\varepsilon^{\beta_1\alpha_2}\varepsilon^{\beta_2\alpha_3}\cdots\varepsilon^{\beta_{L-1}\alpha_L},
\end{equation}
where $\varepsilon^{\alpha\beta}$ is the antisymmetric tensor with
$\varepsilon^{12}=1$. As shown in \cite{AKLT}, the inner product
of two AKLT chains with L sites is
\begin{equation}
\Omega^{\dag\alpha}_{\beta}\cdot\Omega_{\gamma}^{\delta}=\delta^{\alpha}_{\gamma}\delta^{\beta}_{\delta}(3^L-1)/2
+\delta^{\alpha}_{\beta}\delta^{\delta}_{\gamma},
\end{equation}
so the normalization of the ground state with periodic boundary
conditions reads
\begin{equation}
\Omega^{\dag\beta}_{\beta}\cdot\Omega_{\alpha}^{\alpha}=3^L+3
\end{equation}

Next we calculate the spin-spin correlation function between the
site 1 and 2 $\langle S_1^z S_2^z\rangle$ for a chain with L sites
and periodic boundary conditions. Following \cite{AKLT}, the spin
operators acting on our basis for a single spin 1 give rise to
\begin{eqnarray}
S_1^z
\psi_{\alpha\beta_1}&=&(1/2)[(-1)^\alpha+(-1)^{\beta_1}]\psi_{\alpha\beta_1},\\
S_2^z
\psi_{\alpha_2\beta_2}&=&(-1/2)[(-1)^{\alpha_2}+(-1)^{\beta_2}]\psi_{\alpha_2\beta_2}.
\end{eqnarray}
Combining $A(2)-A(6)$, we can find
\begin{equation}
\langle S_1^zS_2^z\rangle=-4(3^{L-2}-1)/(3^L+3).
\end{equation}
We notice that the AKLT chain is isotropic,  so $\langle\mathbf{S}_{\mathbf{_1}%
} \cdot \mathbf{S}_{\mathbf{_2}}\rangle$=3$\langle S_1^z
S_2^z\rangle=-12(3^{L-2}-1)/(3^L+3)$. Now we have proven the fact
that the average value of $\mathbf{S}_{\mathbf{_i} } \cdot
\mathbf{S}_{\mathbf{_i+1}}$ is lower in an AKLT chain with
infinite length than that in a closed AKLT loop with finite
length.
When $L\rightarrow\infty$, we get the value in \cite{AKLT} that $\langle\mathbf{S}_{\mathbf{_i}%
} \cdot \mathbf{S}_{\mathbf{_i+1}}\rangle=-4/3$.


\begin{references}
\bibitem[*]{} Electronic address: schen@aphy.iphy.ac.cn
\bibitem{Misguis}G. Misguish and C. Lhuillier,
 Review article in the book ``{\it Frustrated spin systems}", edited by H.
T. Diep, World Scientific,Singapore (2005), and references
therein.
\bibitem{2}P. Chandra and B. Doucot, Phys. Rev. B {\bf
38}, 9335 (1988).
\bibitem{3} T. Einarsson and H. Johannesson, Phys.
Rev. B {\bf 43}, 5867 (1991).
\bibitem{4}  E. Fradkin, {\it Field
theories of condensed matter systems}. (Addison-Wesley Publishing
company,1991)
\bibitem{5}L. Capriotti and S.
Sorella, Phys. Rev. Lett. {\bf 84}, 3173 (2000).
\bibitem{Read}N.
Read and S. Sachdev, Phys. Rev. B {\bf 42}, 4568 (1990); Phys.
Rev. Lett. {\bf 66}, 1773 (1991).
\bibitem{Capriotti01}L. Capriotti, F.
Becca, A. Parola, and S. Sorella, Phys. Rev.Lett. {\bf 87}, 097201
(2001); Phys. Rev. B {\bf 67}, 212402 (2003).
\bibitem{Sushkov}O. P. Sushkov, J. Oitmaa, and Z. Weihong, Phys. Rev. B {\bf 63}, 104420
(2001).
\bibitem{Cirac}J. J. Garcia-Ripoll, M. A. Martin-Belgado, and J. I. Cirac,
Phys. Rev. Lett. {\bf 93}, 250405 (2004).
\bibitem{7}F. D. M. Haldane, Phys. Rev. Lett. {\bf 61}, 1029 (1988).
\bibitem{Arovas}D. P. Arovas, A. Auerbach, and F. D. M. Haldane, Phys. Rev. Lett. {\bf 60}, 531 (1988).
\bibitem{AKLT}I. Affleck, T. Kennedy, E. H. Lieb, and H. Tasaki, Phys. Rev. Lett. {\bf 59}, 799 (1987); Commun. Math.
Phys. {\bf 115}, 477 (1988).
\bibitem{17}M. den Nijs and K. Rommelse, Phys. Rev. B {\bf 40}, 4709 (1989)
\bibitem{Haldane83}F. D. M. Haldane, Phys. Rev. Lett. {\bf 51}, 605 (1983).
\bibitem{Batista}C. D. Batista and S. A. Trugman, Phys. Rev. Lett. {\bf 93}, 217202 (2004).
\bibitem{Majumdar} C. K. Majumdar and D. K. Ghosh, J. Math. Phys. {\bf 10},
1388(1969); B. S. Shastry and B. Sutherland, Physica B {\bf 108},
1069 (1981).
\bibitem{Chen} S. Chen, H. Buttner and J. Voit, Phys. Rev. B {\bf 67},
054412, (2003); Phys. Rev. Lett. {\bf 87}, 087205, (2001); S. Chen
and B. Han, Eur. Phys. J. B {\bf 31}, 63, (2003).
\bibitem{Nussinov} Z. Nussinov and A. Rosengren, Cond-mat/0504650.
\bibitem{Sen} D. Sen and N. Surendran, Phys. Rev. B {\bf 75},104411
(2007).
\bibitem{Zittartz} A. Klumper, A. Schadschneider, and J. Zittartz, J.
Phys. A {\bf 24}, L955 (1991).
\bibitem{18}H. Tasaki, Phys. Rev. Lett {\bf 66}, 798  (1991).
\bibitem{Wen}X.-G. Wen, Phys.Rev. Lett {\bf 90}, 016803
(2003).
\end{references}
\end{document}